# Invisible Sensors: Simultaneous Sensing and Camouflaging in Multiphysical Fields

*Tianzhi Yang, Xue Bai, Dongliang Gao, Linzhi Wu, Baowen Li, John T. L. Thong, and Cheng-Wei Qiu\**

When a sensor is used to probe a physical field, it intrinsically disturbs surrounding environment, introduces perturbation and unwanted noise in the measurement. A sensor with a larger cross section may, indeed, extract, capture, and measure the data of interest more easily. However, its presence will also generate a stronger perturbation. Therefore, most sensing systems are essentially "visible" and may need to be camouflaged in many applications. Recently, this problem has been addressed and solved by using the scattering-cancellation technique.[1] This technique provides a metamaterial shell to cover the sensor for suppressing the overall scattering while maintaining its receiving ability. However, this technique is only valid for a single physical field, e.g., an invisible electromagnetic sensor or an invisible acoustic detector.[2–4] Consequently, a single-functional sensor is invisible to an acoustic monitoring receiver, but it can be easily detected using a remote thermal imager. Is it possible to create a sensor that is invisible in multiple physical fields while maintaining the same sensing functionality?

This is very challenging, if not impossible, to achieve, even using the concept of metamaterials, which are man-made composites that control waves and energy flux in unprecedented ways, resulting in exotic behaviors that are absent in nature. For example, electromagnetic metamaterials were proposed to manipulate electromagnetic waves and produce an invisibility cloak.[5–7] This pioneering idea motivated a number of significant applications, such as the wave concentrator and rotator.[8–10] Other than the electromagnetic waves, metamaterials have been created to manipulate other waves such as acoustic waves,[11–13] elastic waves,[14,15] magnetostatic fields,[16] and static forces.[17] More recently, metamaterials were presented that control the DC current[18–26] and the heat flux.[27–37] However, these devices were designed to cloak an object in a single physical field.

Advanced and multifunctional metamaterials are highly desirable for most practical applications. More recently, some attempts to cloak an object in multiple physical fields have been made, in particular, the bifunctional thermal-electric invisibility cloak[38] and independent manipulation[39] were proposed. Later, the first experiment was carried out to simultaneously cloak an air cavity in the electric and thermal fields.[40] This sample was fabricated through a sophisticated man-made metamaterial structure with many holes drilled in a silicon plate that were, then, filled with poly(dimethylsiloxane) (PDMS). In our work, we found that natural materials with simple structure can also simultaneously manipulate multiphysical fields. We fabricated a device that acted as a "mask" for both thermal and electric fields and behaved as a multifunctional invisible sensor.

To date, the theory of "cloaking a sensor" is only valid for a single physical field.[1] In this study, we present the first invisible sensor theory for static multiphysical-field. This multiphysical invisible sensor has three features that distinguish it from conventional DC and thermal metamaterial devices,[18–39] especially different from the bifunctional cloak for an air cavity.[40] First, we allow the sensor to "see through and behind" the cloaked region in multiphysical fields. As a result, the sensor is invisible and receives proportional incoming signals at the same time, and it is able to "open its eyes" behind the cloak to receive information from the outside multiphysical

Dr. T. Z. Yang
Department of Astronautics
Shenyang Aerospace University
Shenyang 110136, China
Dr. T. Z. Yang, X. Bai, Prof. J. T. L. Thong,
Prof. C.-W. Qiu
Department of Electrical and Computer Engineering
National University of Singapore
Kent Ridge 117583, Republic of Singapore
E-mail: eleqc@nus.edu.sg
X. Bai, Prof. B. Li,[+] Prof. J. T. L. Thong, Prof. C.-W. Qiu
NUS Graduate School for Integrative Sciences and Engineering
National University of Singapore
Kent Ridge 117456, Republic of Singapore
Dr. D. L. Gao
College of Physics
Optoelectronics and Energy & Collaborative Innovation Center
of Suzhou Nano Science and Technology
Soochow University, Suzhou 215006, China
Prof. L. Z. Wu
Center for Composite Materials
Harbin Institute of Technology
Harbin 150001, China
Prof. B. Li
Department of Physics and Centre for Advanced 2D Materials
National University of Singapore
Kent Ridge 117546, Republic of Singapore
Prof. B. Li
Center for Phononics and Thermal Energy Science
School of Physics Science and Engineering
Tongji University
Shanghai 200092, China
[+]Present address: Department of Mechanical Engineering, University of Colorado, Boulder, CO, 80309-0427, USA.

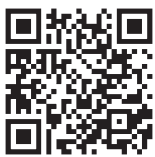

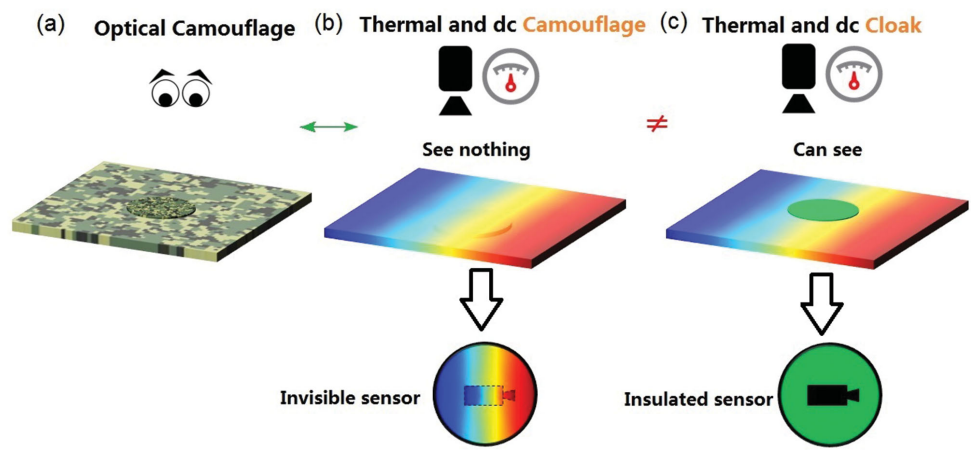

**Figure 1.** Illustration of analog of optical camouflage, our bifunctional camouflage, and conventional bifunctional cloak. a) Illustration of the optical camouflage. b) Thermal and DC camouflage, for both in-plane and out-of-plane detection. c) Conventional thermal and DC cloak for in-plane invisibility.

world. By contrast, a conventional multiphysical cloaking device isolates and prevents any signal from propagating/entering the cloaked region. However, for a 2D conventional cloak, as shown in **Figure 1**, the cloaked object is still visible for out-of-plane detection using the thermal imaging and DC current scanning apparatus. By contrast, we make the sensor truly invisible to both in- and out-of-plane remote observers. Second, our invisible sensor is bifunctional, and is undetectable both in thermal and electric fields. This is a more advanced technique because previous "cloaking a sensor" techniques were only valid in a single physical field.[1–4] To this day, no theoretical or experimental works have been proposed that cloak or camouflage a sensor in a multiphysical environment. We believe this is a step toward cloaking a sensor in a realistic environment, without it being detected by different types of remote observers. Third, it is believed that the thermal metamaterials can also manipulate DC electric current simultaneously, because thermal conductivity is proportional to electric conductivity for some materials, such as metals. However, it is extremely difficult to find several types of natural materials with exactly the same proportions to build an ideal metamaterial device, as shown in **Table 1**. Here, we propose a tuning method for utilizing natural materials with nonproportional thermal and electric conductivities, which was considered impossible to accomplish.[39]

A schematic diagram of our dual invisible sensor concept is shown in Figure 1. Figure 1a shows a classical bio-inspired camouflage technique in visible region, in which a sensor is camouflaged through mimicking the background color. This surface coloration camouflage method can be extended from visible region to multiple physical fields, as shown in Figure 1b. Here, we stress that our camouflage technique is realized through mimicking the surrounding field, instead of cloaking the sensor using an insulator. Our scheme is single-blind, which makes the outside observer unable to detect the sensor, while the sensor can see outside. This feature is drastically different from the conventional bifunctional cloaking technique,[40,41] in which the cloaked region is fully isolated, as shown in Figure 1c (green region). The conventional cloaking devices block the incoming heat flux, and the heat signature of the cloaked region looks too "dark" and remarkably different from the surrounding environment in the thermal image. This makes the sensor detectable for a remote out-of-plane observer, though the sensor remains undetectable in-plane. Consequently, the conventional cloak is not suitable for cloaking a sensor.

In a typical situation, a sensor may be exposed to a multiphysical environment. To obtain an ideal invisible sensor, we began our studies by covering the sensor with a thin shell (with thickness $b-a$). The sensor (with radius $a$) is embedded in a background plate and is subjected to the uniform thermal and electric fields at both ends. Such a built-in sensor can be found in practical applications, for example, in a semiconductor wafer-processing hotplate. Due to the discontinuity of material conductivities, the presence of the bare sensor disturbs the surrounding thermal and electric signatures. To realize the bifunctional invisible sensing device, we use a camouflage shell that is expected to drastically reduce the perturbation of heat flux and electric current simultaneously. To camouflage such a sensor in multiphysical fields, we need to consider the electric and thermal conduction equations simultaneously:

$$\nabla \cdot (\kappa \nabla T) = 0, \nabla \cdot (\sigma \nabla V) = 0 \quad (1)$$

where $\kappa$ and $T$ denote the thermal conductivity and temperature, $\sigma$ and $V$ denote the electrical conductivity and electrical potential. These two equations can be rewritten in a general form, given in Equation (2):

**Table 1.** Commonly used constituted nature materials in previous thermal metamaterials.

| Materials | $\sigma$ [S m$^{-1}$] | $\kappa$ [W m$^{-1}$ K$^{-1}$] | $\eta$ ($\sigma/\kappa$) | $\overline{\eta}$ (Relative magnitude to copper) |
|---|---|---|---|---|
| Copper | $5.9 \times 10^7$ | 400 | $1.48 \times 10^5$ | 1 |
| Aluminum | $3.7 \times 10^7$ | 220 | $1.68 \times 10^5$ | 1.14 |
| Lead | $4.8 \times 10^6$ | 35 | $1.37 \times 10^5$ | 0.93 |
| Tungsten | $1.89 \times 10^7$ | 173 | $1.1 \times 10^5$ | 0.75 |
| Stainless steel (436) | $1.43 \times 10^6$ | 30 | $4.77 \times 10^4$ | 0.32 |
| Magnesium alloy | $6.99 \times 10^6$ | 72.7 | $9.6 \times 10^4$ | 0.65 |

$$\nabla \cdot (\lambda \nabla \phi) = 0 \tag{2}$$

The general solutions $\phi_i$ of Equation (2) can be expressed as:

$$\phi_i = \sum_{n=1}^{\infty} [A_m^i r^m + B_m^i r^{-m}] \cos m\theta \tag{3}$$

where $A_m^i$ and $B_m^i$ are constants to be determined by boundary conditions. ($r$, $\theta$) represent cylindrical coordinates. $\phi_i$ denotes the temperature and potential in different regions: $i = 1$ for the inner camouflaged sensor, $i = 2$ for the shell and $i = 3$ for the exterior background. To cloak the sensor without affecting its capability to receive the incoming signals, the material parameters of the concealed sensor, shell and background can be derived from boundary conditions:

$$\phi_i|_{r=a,b} = \phi_{i+1}|_{r=a,b} \quad \lambda_i \frac{\partial \phi_i}{\partial r}\bigg|_{r=a,b} = \lambda_{i+1} \frac{\partial \phi_{i+1}}{\partial r}\bigg|_{r=a,b} \tag{4}$$

$\lambda_i$ ($i = 1, 2, 3$) denotes the electric or thermal conductivities of the sensor ($r \leq a$), shell ($a < r \leq b$), and the external region ($r \geq b$), respectively. If we force the external signals to enter the shell, we have $\lambda_3 = \lambda_b$, where $\lambda_b$ is the thermal or electric conductivity of the background. The substitution of Equation (3) for Equation (4) yields:

$$b = a \sqrt{\frac{(\lambda_2 - \lambda_1)(\lambda_2 + \lambda_b)}{(\lambda_2 + \lambda_1)(\lambda_2 - \lambda_b)}} \tag{5}$$

The general form Equation (5) can be rewritten for thermal and electric fields, respectively:

$$b_{thermal} = a \sqrt{\frac{(\kappa_2 - \kappa_1)(\kappa_2 + \kappa_b)}{(\kappa_2 + \kappa_1)(\kappa_2 - \kappa_b)}}, \quad b_{elec} = a \sqrt{\frac{(\sigma_2 - \sigma_1)(\sigma_2 + \sigma_b)}{(\sigma_2 + \sigma_1)(\sigma_2 - \sigma_b)}} \tag{6}$$

where $b_{thermal}$ and $b_{elec}$ are the required shell radii for thermal and electric fields. The thickness of thermal and electric shells are defined as $b_{thermal}$-$a$ and $b_{elec}$-$a$, respectively. To realize a bifunctional device, it is necessary to match $b_{thermal} = b_{elec}$ for a given sensor radius $a$, which means that the thermal and electric shells have the same thickness. Therefore, based on Equation (6), we have the relation:

$$\frac{(\sigma_2 - \sigma_1)(\sigma_2 + \sigma_b)}{(\sigma_2 + \sigma_1)(\sigma_2 - \sigma_b)} = \frac{(\kappa_2 - \kappa_1)(\kappa_2 + \kappa_b)}{(\kappa_2 + \kappa_1)(\kappa_2 - \kappa_b)} \tag{7}$$

Here we introduce a parameter $\eta_i = \sigma_i / \kappa_i (i = 1, 2, b)$. Note that we did not use $\eta_1 = \eta_2 = \eta_b$ to fulfill Equation (7). It is well known that natural materials exhibit very similar/approximate ratios between the electric and thermal conductivities based on the empirical Wiedemann–Franz law, while the constant varies for different materials. However, as shown in Table 1, we list nature materials that were commonly used for thermal metamaterials in previous references. We found that their ratio are not exactly equal: $\eta_1 \neq \eta_2 \neq \eta_3... \neq \eta_i$, and consequently, the condition (7) cannot be exactly fulfilled for three types of natural materials. This is the essential reason that most existing metamaterials are only valid for manipulating one physical field, thus serving only a single-function application.

Because of this inherent difficulty, we note that the previous publication on multiphysical cloak[40] resorts to complex metamaterials, by drilling holes to manipulate the averaged effective thermal and DC conductivities to fulfill condition (7). That case is, actually, a subset of our model (corresponding to $\sigma_1 = \kappa_1 = 0$), for which a special solution was imposed as $\eta_2 = \eta_b$. However, two randomly selected nature materials do NOT fulfill this condition, which is easily supported by choosing two types of normal bulk materials from Table 1 for a test. Therefore, Ma et al.[40] experimentally demonstrated the multiphysical cloak using a porous metamaterial structure to achieve $\eta_2 = \eta_b$. However, we found that it is not necessary to pursue $\eta_1 : \eta_2 : \eta_b = 1:1:1$ in this bifunctional device. In this work, we show that some natural materials can fulfill Equation (7) without resorting to complex decorations. The use of undecorated materials throughout has previously been considered to be impossible,[39] because of the belief that the bulk properties of naturally occurring materials are intrinsic and lack flexibility and variation to satisfy Equation (7). Without resorting to metamaterials, we counter-intuitively overcome this utmost and intriguing above-mentioned problem. We use three types of naturally undecorated materials to experimentally demonstrate a multiphysical invisible sensor that functions in thermal and electric fields simultaneously. It is admitted that one needs to judiciously look up the material properties to match the geometrical specifications in Equation (7) for both thermal and DC fields. However, this does not guarantee that there is always a solution (for some other cases there may have been several solutions). In this study, the natural materials we used were stainless steel, copper and magnesium alloy (The parameters of these materials are listed in Table 1). Equation (7) shows that $b_{thermal}$ and $b_{elec}$ rely on the summation and differentiation of the thermal and electric conductivities, instead of maintaining the same ratio $\eta_i$. Therefore, we fix the radius of the sensor $a$ and tune the outer radius $b$ using different naturally available materials (with different thermal and electrical conductivities) to achieve $b_{thermal} \approx b_{elec}$ using three types of natural materials with relative ratios of $\eta_1 : \eta_2 : \eta_b \approx 1:3.1:2$, instead of the special case of 1:1:1. This provides a plausible route for designing a multi-functional invisible sensor using natural materials.

As an example, we consider a stainless steel sensor with a coated tungsten shell embedded in a lead plate. We plot the required radius $b$ in **Figure 2**a both in the thermal and electric fields. The solid black and green dashed lines correspond to the calculated radius $b_{elec}$ and $b_{thermal}$, respectively. It can be seen that, for a fixed sensor radius $a$, the calculated radius $b_{elec} > b_{thermal}$, which implies that the tungsten shell is *not* bifunctional. For example, here we chose the sensor radius as $a = 15$ mm and obtained $b_{thermal} = 15.46$ mm and $b_{elec} = 18.03$ mm. If we choose $b_{thermal}$ as the radius of the shell, the corresponding numerical simulations based on the finite element method (FEM) are shown Figure 2c, in which the red arrows denote the heat flux and the black lines denote the isopotential lines. It can be seen that the hear flux propagation is not disturbed and can penetrate the shell. By contrast, because of the size mismatching, the sensor strongly distorts the DC current in the electric field, creating an obvious shadow near the sensor. As the electric current gets closer to the sensor, the potential profile becomes more distorted. As a result, the sensor is easily detected and localized, and thus behaves as a single-functional invisible sensor.

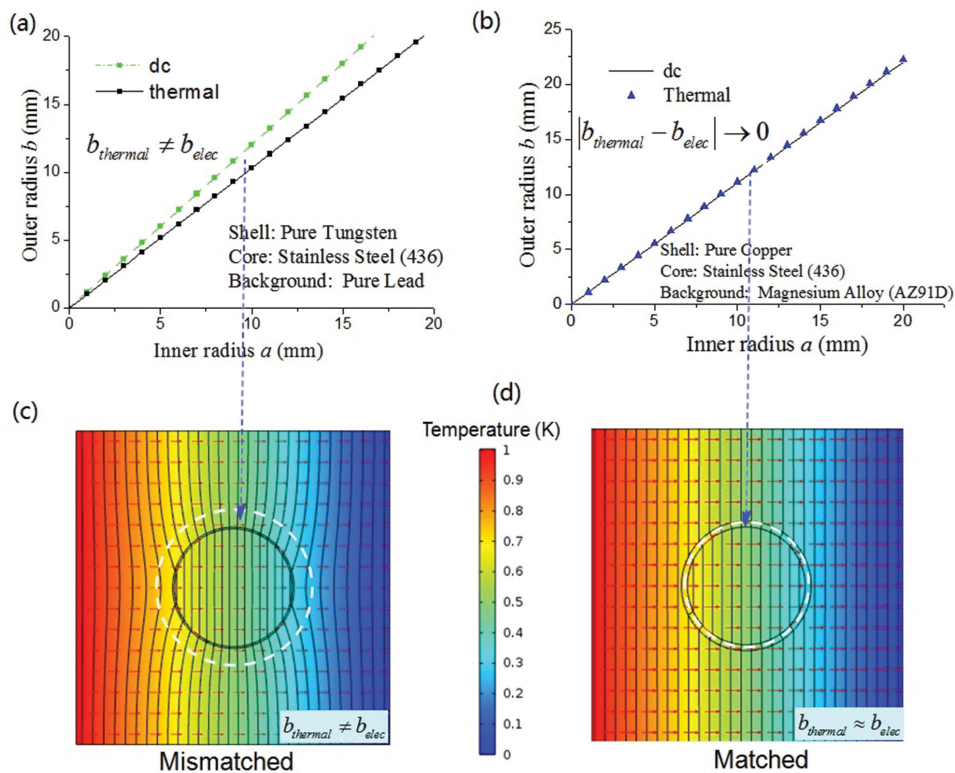

**Figure 2.** Design for the invisible sensor in thermal and electric fields. a) A mismatched case for camouflaging a sensor made of stainless steel coated with a tungsten shell, inserting in a Lead plate. b) A matched case for camouflaging a sensor made of stainless steel coated with a copper shell, inserting in a magnesium alloy plate. c) The corresponding FEM simulation for the mismatched case. The black circle denotes the required radius $b_{elec}$, and the white dashed circle denotes the required radius $b_{thermal}$. d) The corresponding FEM simulation for the matched case. The black circle denotes the required radius $b_{elec}$, and the white dashed circle denotes the required radius $b_{thermal}$.

As another example, we choose copper and magnesium alloy as materials for the sensor and the background, respectively. As shown in Figure 2b, the solid black line and the blue triangular points correspond to the calculated $b_{thermal}$ and $b_{elec}$, respectively. In this case, the two lines agree well as radius $a$ increases, which implies that $b_{thermal} \approx b_{elec}$, even for a very large sensor. This makes it possible for the unique shell to camouflage a sensor not only in the thermal but also in the electric fields: by matching the material conductivities and the structural size, and one can thus create a unique shell with dual minimal perturbation. The corresponding FEM simulation is shown in Figure 2d, the heat flux and DC current are not disturbed by the coated sensor. It demonstrates that the copper shell exhibits the dual-field camouflage performance and serves as a bifunctional invisible sensing device. These three types of natural materials have very different relative ratios of $\eta_1 : \eta_2 : \eta_b \approx 1 : 3.1 : 2$. It should be emphasized that, this tuning method is valid over a wide range of different natural materials.

As a proof of a concept, we performed more detailed simulations together with experiments to validate our device. The experimental setup, materials, and the structural parameters are shown in **Figure 3**. As shown in **Figure 4**a1, the vertical lines represent the isothermal contour. When the sensor (with radius $a = 15$ mm) is wrapped with the copper shell (with a radius $b = 16.52$ mm to achieve $b_{thermal} \approx b_{elec}$), the sensor disappears in the thermal image and smoothly merges into the background. For our fabricated sample, the thickness of the shell is 1.52 mm, the heat flux penetrates the shell and the sensor receives proportional incoming signal from outside. To design an ideal shell, the thickness of the shell is a crucial parameter for perfect camouflage performance. The "sensing" function performs well because the sensor can probe the original signal strength (temperature or potential) at its location, proportionally depending on the thickness of the shell based on Equation (6). One could see that, the thinner the shell is, the lesser the disturbance to the external field is. Fortunately, ultrathin shell is not difficult for the current technologies, and an extremely small thickness for the shell could enable a nearly perfect camouflage performance, for example, at a shell thickness of 100 μm the disturbance to the external field gradient is negligible. The natural materials of high conductivities, e.g., copper and silver, are good candidates for achieving this goal, when they are properly codesigned with the material and geometry of the core. The properly designed multiphysical shell renders the sensor completely invisible and enables the sensor to measure the external field because the coating shell eliminates surrounding thermal field distortions. As a reference sample, a bare sensor was simulated in Figure 4a2. As a poor conductor, the bare sensor strongly attracts and bends the isothermal lines toward itself. As a result, the presence of the sensor distorts the surrounding thermal field and exhibits an obvious heat signature in thermal imaging. The sensor can be easily captured and observed using an out-of-plane observer, such as an infrared camera. Moreover, the sensor receives an inaccurate temperature signal in the distorted field.

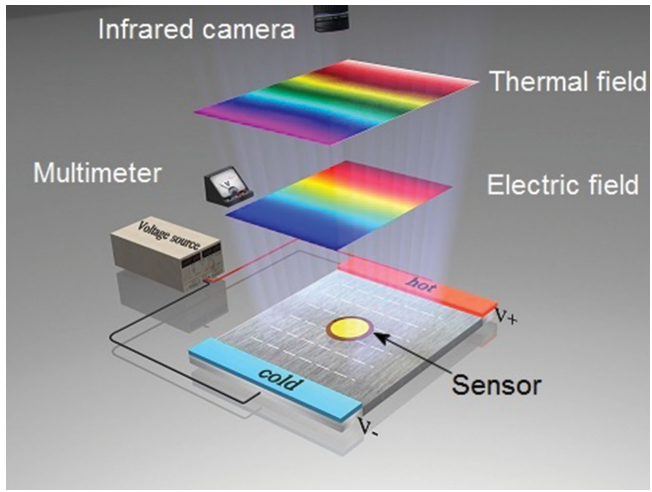

**Figure 3.** Schematic illustration of the 2D simultaneous thermal and electric invisible sensor with unprocessed materials. In the experimental setup, the sensor, shell and the background are made of pure copper, stainless steel (ASTM 436) and magnesium alloy (AZ91D), respectively. The thermal and electric fields are applied on both ends of the sample. The inset shows the fabricated sample in greater detail. The geometrical parameters are $a = 15$ mm, $b = 16.52$ mm.

Figure 4b1 shows the corresponding experimental results of the fully camouflaged sensor in the thermal field. As expected, the thermal signature of the sensor disappears entirely. Without marking the sensor position, we cannot observe and localize the sensor using an infrared camera. This camouflage result is remarkably different from the previous DC[18–26] and thermal cloaks,[27–37,40,42] in which the cloaked object looks much cooler in the thermal or electric signature. In our work, we make the heat signature of the sensor emerge into the background. For a comparison, the strongly perturbed heat diffusion was observed for the bare sensor, as shown Figure 4b2. The isothermal lines are obviously bent as heat flux propagates through the bare sensor. The bare sensor automatically appears in the observed thermal signature, thus it is easy to see the shape, size, and position of the sensor. Moreover, the sensor cannot measure the correct temperature data due to the disorder of isothermal lines. Remarkably, the experimental results in Figure 4 are consistent with the numerical simulations. (A detailed experimental recording movie can be found in the Supporting Information.[43]) This movie clearly shows the advances of the current "invisible sensor." It is noted that our dual-field camouflaging theory is steady-state based. It can be seen from this movie that our device exhibits very good camouflage performance in steady-state. However, one can still see very slight sensor signature and the shell a bit in the experiment during the transient process toward the steady state. (For more details theoretical and experimental analysis of the transient behavior, please refer to ref. [44].) This transient camouflage performance can also be enhanced using an ultra-thin shell, for example, 100 μm. The thermal camera, even with very high resolution, would find it difficult to detect both the sensor and the shell.

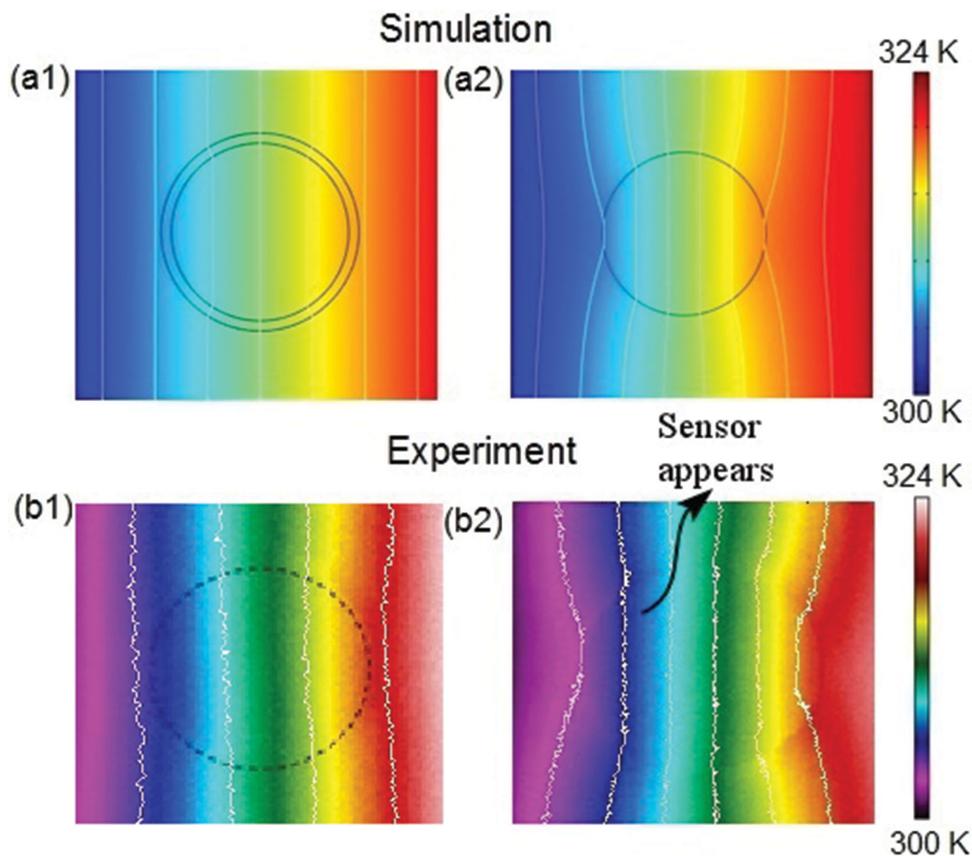

**Figure 4.** Calculated and measured thermal field distribution. a1) Calculated thermal field distribution of a camouflaged sensor. a2) Calculated thermal field distribution of a bare sensor. b1) Measured thermal field distribution of a camouflaged sensor. The sensor, marked by the black dotted circle, is completely camouflaged in the surrounding environment. The white solid lines denote the observed isothermal lines. b2) Measured thermal field distribution of the bare sensor. Without marking the position of sensor, a clear visible outline of the sensor is observed. The bare sensor can thus be easily detected using an infrared camera.

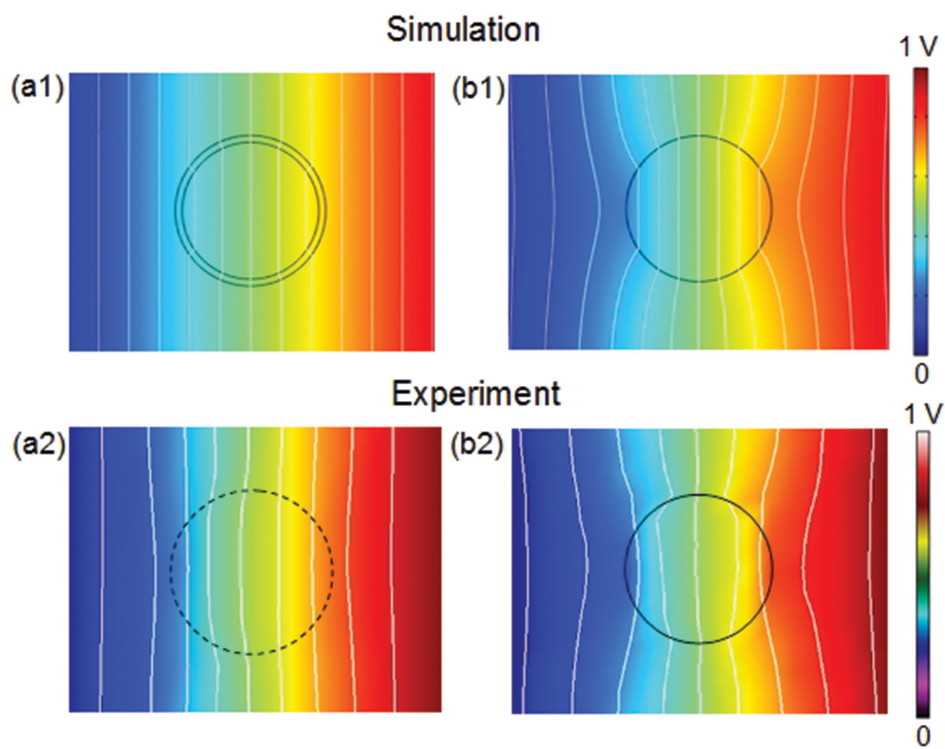

**Figure 5.** Calculated and measured electric field distribution. a1) Calculated electric field distribution of the camouflaged/invisible sensor. a2) Calculated electric field distribution of a bare sensor. b1) Measured electric field distribution of the camouflaged/invisible sensor. The sensor, marked by the black dotted circle, is completely camouflaged in the surrounding environment. b2) Measured electric field distribution of the bare sensor.

We proceeded to validate the camouflage performance in the electric field. **Figure 5**a1 shows the simulated potential contour of the coated sample, in which the solid lines refer to the isopotential lines. It can be seen that the isopotential lines are straight and parallel. The electric current smoothly enters and passes through the sensor without deviating the original trajectories, which renders the sensor entirely invisible. As a result, the sensor is concealed and cannot be detected or localized. By contrast, as shown in Figure 5a2, the simulated potential lines are strongly disturbed by the bare sensor. In the vicinity of the sensor, the isopotential lines are strongly bent. Therefore, the sensor can be easily detected using an electric impedance scanner.

Figure 5b1 shows the corresponding experimental result in the electric field. The shell can be seen to eliminate the "shadow" in the near-field, and the sensor smoothly fits the outside electric field, and exhibits a good camouflage performance. The electric current perturbation is greatly suppressed around the sensor. The presence of the coated sensor does not disturb the surrounding electric field. Therefore, the sensor can receive the proper incoming DC current. In comparison, the bare sensor (as shown in Figure 5b2) strongly distorts the potential distribution. As the electric current gets closer to the sensor, the potential profile becomes more distorted and creates an obvious shadow. As a result, the sensor can be easily detected and localized. All these measurements in the electric field are in good agreements with our theoretical predictions in Figure 5a1, a2.

In conclusion, we have theoretically and experimentally presented a 2D, multiphysical invisible sensor made of the isotropic and naturally available materials. We show that it is possible to simultaneously suppress multi-field perturbation of a sensor without blocking the incoming signals. Thus this invisible sensor can simultaneously avoid being noticed by an infrared thermal imaging camera and electrical impedance scanning. The presented bifunctional camouflage shell is very thin, isotropic, homogeneous, and is facile to fabricate in practice. This recipe can be readily extended to the three-dimensional cases.

### Experimental Section

For the experimental verification, samples were fabricated using an electrical discharge machining (EDM) process. The shell and the sensor were tightly assembled to exclude air at the interface. The steady-state temperature measurement was taken after 10 min. The temperature profile was using two infrared cameras (FLIR i60 and FLIR T620). A constant temperature gradient was imposed across the samples by maintaining the hot and cold sides. The hot side was maintained using a heater, and the cold side was an ice-water mixture, held at 273 K. The top and bottom edges were insulated.

To quantitatively test the performance of the camouflage in the electric field, the potential distribution was probed using a Keithley 2000 multimeter. The electric field measurement was taken at room temperature (20 °C). Because the samples are good electric conductors, a powerful DC current source was used to pass a large electric current (8A) through the samples to obtain the noise-free voltage data. A constant voltage $V_{total}$ was imposed on the left and right boundaries of the entire sample. The top and bottom boundaries were physically insulated. To obtain stable and accurate data, the values $V_i$ at each node were measured and then normalized to the total potential $V_{total}$. Each node was measured five times and the average value was processed in the MATLAB environment.

### Acknowledgements

T.Z.Y. and X.B. contributed equally to this work. T.Z.Y. and D.L.G. acknowledge support from the National Science Foundation of China under Grant No. 11202140 and 11504252, Kong-Tian Scholarship from school of Aerospace Engineering in Shenyang Aerospace University and Program for Liaoning Excellent Talents in University (No. LJQ2013020). L.Z.W. acknowledges support from the National Science Foundation of China under Grant No. 11432004. The authors also gratefully acknowledge the assistance of Prof. Baile Zhang and Dr. Hongyi Xu in infrared imaging.

Received: May 26, 2015
Revised: August 7, 2015
Published online: